# Spin transport and spin conversion in compound semiconductor with non-negligible spin-orbit interaction

Akiyori Yamamoto[1], Yuichiro Ando[1,2], Teruya Shinjo[2], Tetsuya Uemura[3], and Masashi Shiraishi[1,2,#]

1. Graduate School of Engineering Science, Osaka Univ., Toyonaka 560-8531, Japan.
2. Department of Electronic Science and Engineering, Kyoto Univ., Kyoto 615-8510, Japan.
3. Graduate School of Information Science and Technology, Hokkaido University, Sapporo 060-0814, Japan

# Corresponding author: Masashi Shiraishi (mshiraishi@kuee.kyoto-u.ac.jp)

Abstract

   A quantitative investigation of spin-pumping-induced spin-transport in $n$-GaAs was conducted at room temperature (RT). GaAs has a non-negligible spin orbit interaction, so that electromotive force due to the inverse spin Hall effect (ISHE) of GaAs contributed to the electromotive force detected with a platinum (Pt) spin detector. The electromotive force detected by the Pt spin detector had opposite polarity to that measured with a $Ni_{80}Fe_{20}$/GaAs bilayer due to the opposite direction of spin current flow, which demonstrates successful spin transport in the $n$-GaAs channel. A two-dimensional spin-diffusion model that considers the ISHE in the $n$-GaAs channel reveals an accurate spin diffusion length of $\lambda_s = 1.09$ μm in $n$-GaAs ($N_{Si} = 4\times10^{16}$ cm$^{-3}$) at RT, which is approximately half that estimated by the conventional model.

Spin-pumping-induced spin transport (SPIST), of which a typical device consists of a ferromagnetic spin injector and a nonmagnetic spin detector separated by a nonmagnetic channel, enables a direct estimation of the spin transport properties of the channel.[1-4] Measurement of room temperature (RT) spin transport by the SPIST method has been so far reported for $p$-type silicon, graphene, aluminum, and a conductive polymer.[1-4] In SPIST, the transfer of spin angular momentum due to the spin pumping generates a spin accumulation in the channel adjacent to the spin injector, which results in the flow of a pure spin current in the channel.[5-8] The transported spin current is electrically detected via the inverse spin Hall effect (ISHE) of the spin detector, in which heavy elements such as platinum (Pt) and palladium (Pd) are often employed.[9-15] Since magnitude of electromotive force due to the ISHE is proportional to the length of the spin injector and detector, spin detection based on the ISHE enables enhancement of the spin detection sensitivity only by changing the device size, which is a significant advantage over spin accumulation voltage measurements.[16-18] RT spin transport in a $p$-type Si has so far only been realized by means of SPIST, due to the difficulty of spin transport caused by a short spin lifetime and small diffusion constant.[1,19] Although SPIST is expected to be applicable to the investigation of spin transport properties in a wide variety of materials, it becomes complicated when the spin transport occurs by two- or three-dimensional diffusion and the spin channel has a considerable spin-orbit interaction (SOI), because the channel also generates an electromotive force. In this case, there is a requirement to distinguish the effective electromotive force in the spin detector from that in the channel.

In this study, we focus on $n$-GaAs because it has a non-negligible SOI (spin Hall angle, $\theta_{SHE}$ = 0.007) and a considerable spin diffusion length of $\lambda_s$ = 1.6 μm at 205 K (doping concentration $N_{Si}$ = 5×10$^{16}$ cm$^{-3}$) [20-30] i.e., this is an appropriate material platform to develop an understanding of the interplay between the electromotive force from a semiconductor channel and that from a heavy-metal spin-detector. Comparison of the electromotive force obtained in a lateral SPIST device with that in a conventional spin injector/detector bilayer structure demonstrates successful spin transport in an $n$-type GaAs channel at RT. Theoretical analysis based on the two-dimensional spin diffusion model reveals that a spin current comparable to that in the spin detector flows in the $n$-GaAs channel along the normal direction of the channel plane of the SPIST device. An accurate value of $\lambda_s$ in $n$-GaAs is estimated by considering the electromotive force in both the GaAs channel and the Pt spin detector. This study allows the establishment of a new approach to the investigation of SPIST in condensed matter, in which the ISHE-induced conversion of a pure spin current is not negligible.

Figure 1(a) shows a schematic illustration of a $Ni_{80}Fe_{20}$ (Py)/$n$-GaAs/Pt lateral device. A GaAs channel consisting of 15 nm thick $n^+$-GaAs ($N_{Si}$ = 5×10$^{18}$ cm$^{-3}$)/15 nm thick $n^+ \rightarrow n$-GaAs transition layer/700 nm thick $n$-GaAs ($N_{Si}$ = 4×10$^{16}$ cm$^{-3}$)/250 nm thick $i$-GaAs was grown on a non-doped GaAs(001) substrate by molecular beam epitaxy at 590 °C. The $n^+$-GaAs layer allows highly efficient spin pumping due to a thin Schottky barrier (ca. 10 nm).[26,31,32] Electrical spin injection and detection using the $n^+$-GaAs/$n$-GaAs structure has been demonstrated in nonlocal voltage measurements, which guarantees a considerable $\lambda_s$ in the $n$-GaAs channel.[27-29] Spin transport along the normal direction of the channel plane is expected due to the thick $n$-GaAs channel, which generates an electromotive force along the ±y direction caused by the ISHE in the GaAs channel. After growth of the $n^+$-GaAs/$n$-GaAs layers, the substrate was exposed to air for fabrication of a Py spin injector and a Pt spin detector. A 5 nm thick Pt layer were formed by means of the lift-off process with electron beam (EB) lithography and EB deposition system. After fabrication of the Pt layer, the substrate was cleaned by using acetone and 2-propanol, and then the Py layer was fabricated in the same manner as fabrication of the Pt layer. The length $L_y$ and width $L_x$ of the Py electrode were 500 and 100 μm, respectively. The distance between the Py spin injector and the Pt spin detector was 1.1 μm, which was measured by atomic force microscope. The $n^+$-GaAs layer between the Py and Pt electrodes was etched with a mixed aqueous solution of phosphoric acid and hydrogen peroxide ($H_3PO_4$:$H_2O_2$:$H_2O$ = 4:1:315 v/v). The direction of spin transport was the GaAs<110> direction. Two contacts for measurement of the ISHE were attached to each edge of the Pt electrode with Ag paste.

For the spin pumping measurement, a sample was placed near the center of a $TE_{102}$ cavity of an electron spin resonance (ESR) system (Bruker EMX10/12, microwave frequency: 9.6 GHz), where the rf electric and magnetic field components were at a minimum and maximum, respectively. An external static magnetic field $H$ was applied at an angle of $\theta_H$, as shown in Fig. 1(a). The DC electromotive force $V_{EMF}$ was measured using a nanovoltmeter (Keithley 2400). All measurements were performed at RT.

Figure 1(b) shows the ferromagnetic resonance (FMR) spectra $dI(H)/dH$ as a function of $H$-$H_{FMR}$ for the Py/$n$-GaAs/Pt device measured at $\theta_H$ = 0, 90, and 180°, where $I$ and $H_{FMR}$ are the microwave power absorption intensity and the FMR field, respectively. The microwave power $P_{MW}$ was 200 mW. Clear FMR spectra were observed for all $\theta_H$. Figure 1(c) shows $V_{EMF}$ as a function of $H$-$H_{FMR}$. At $\theta_H$ = 0 and 180°, clear changes in $V_{EMF}$ were observed around $H$-$H_{FMR}$ = 0 mT. The $V_{EMF}$ signals were analyzed using a deconvoluted fitting function with independent contributions from the ISHE (symmetrical Lorentzian curve centered on $H_{FMR}$) and the anomalous Hall effect (AHE;

asymmetrical curve) as follows:[9]

$$V_{\text{EMF}} = V_{\text{ISHE}} \frac{\Gamma^2}{(H-H_{\text{FMR}})^2 + \Gamma^2} + V_{\text{AHE}} \frac{-2\Gamma(H-H_{\text{FMR}})}{(H-H_{\text{FMR}})^2 + \Gamma^2} + aH + b, \quad (1)$$

where $V_{\text{ISHE}}$, $V_{\text{AHE}}$, and $\Gamma$ are the amplitudes of electromotive force for the ISHE and the AHE, and a damping constant, respectively. a and b represent correction coefficients. From Eq. (1), $V_{\text{ISHE}}$ were determined to be -0.97, 1.46 µV at $\theta_H$ = 0 and 180°, respectively. Note that the polarity reversal was observed for $V_{\text{ISHE}}$ when $\theta_H$ was changed from 0 to 180° and $V_{\text{ISHE}}$ at $\theta_H$ = 90° is an order of magnitude smaller than those at $\theta_H$ = 0 and 180°. Such behavior is consistent with the theoretically predicted $V_{\text{ISHE}}$, which is expressed as $V_{\text{ISHE}} \propto J_C \propto J_S \times \sigma$, where $J_C$, $J_S$, and $\sigma$ are the charge current density, the spin current density, and the spin polarized vector, respectively.[9] The sign of $V_{\text{ISHE}}$ indicates that the spin current in the Pt electrode flows toward the +z direction, i.e., the pure spin current flows from the GaAs channel to the Pt detector. The difference in the magnitude of the $V_{\text{ISHE}}$, $|V_{\text{ISHE}}|$, between $\theta_H$ = 0 and 180° is mainly due to the electromotive force caused by the temperature increase of the sample, which has also been reported previously.[33] Therefore, an accurate value of $|V_{\text{ISHE}}|$ is obtained from following equation:

$$|V_{\text{ISHE}}| = \frac{|V_{\text{ISHE}}(\theta_H = 0°) - V_{\text{ISHE}}(\theta_H = 180°)|}{2} \quad (2)$$

$|V_{\text{ISHE}}|$ normalized according to the Py electrode length, $|V_{\text{ISHE}}|/L_y$, was estimated to be 2.42 µV mm$^{-1}$. Figure 1(d) shows the dependence of $|V_{\text{ISHE}}|$ on $P_{\text{MW}}$; $|V_{\text{ISHE}}|$ was proportional to $P_{\text{MW}}$, which is also consistent with the theoretical expectation of spin pumping.[10]

The GaAs channel has a considerable SOI; therefore, unwanted $V_{\text{ISHE}}$ can be generated in the GaAs channel. To estimate $V_{\text{ISHE}}$ generated in the GaAs channel, the electromotive force between both edges of the Py electrode in a Py/GaAs bilayer structure was measured, as shown in Fig. 2(a). Figures 2(b) and 2(c) show the respective FMR spectra and $V_{\text{EMF}}$-$H$ curves measured at $\theta_H$ = 0, 90, and 180°. The deconvoluted curves in Fig. 2(c) show clear $V_{\text{ISHE}}$ signals at $\theta_H$ = 0 and 180°, and no $V_{\text{ISHE}}$ signals were observed at $\theta_H$ = 90°. $|V_{\text{ISHE}}|/L_y$ was estimated to be 2.21 µV mm$^{-1}$, which is comparable with that of the SPIST device and indicates that $V_{\text{ISHE}}$ from the GaAs channel should be taken into account for estimation of the spin transport properties. It should be noted that the polarity of $V_{\text{ISHE}}$ is opposite, as shown in Fig. 1(c) and Fig. 2(c). The positive $V_{\text{ISHE}}$ at $\theta_H$ = 0° shows that the direction of the spin current is -z because *n*-GaAs has a positive spin Hall angle. Therefore, the polarity reverse of $V_{\text{ISHE}}$ between the bilayer structure and the SPIST device demonstrates successful spin transport in the *n*-GaAs channel.

Next, $\lambda_s$ in the GaAs channel was estimated. A considerable $V_{ISHE}$ was detected in the bilayer system, which indicates that the spin current in the GaAs channel flows along the ±z direction. Therefore, a two-dimensional flow of the spin current in the x-z plane of Fig. 1(a) should be considered. We assumed that there was no spin current along the y direction, because uniform spin density along the y direction can be realized according to the geometry of the Py and Pt electrodes. Effects of over etching of the $n$-GaAs layer (15 nm in thick) due to the wet etching process and formation of a depletion layer (150 nm in width) at the $n$-GaAs surface after etching of the $n^+$-GaAs layer were also taken into account. Therefore, the injected spins diffuse vertically down through $n^+$-GaAs and 150-nm of $n$-GaAs before lateral diffusion, as shown in Fig. 3(a).

Firstly the spin current density generated at the Py/GaAs interface $J_s^0$, was estimated from the width of the FMR spectrum. $J_s^0$ is expressed as

$$J_S^0 = \frac{g_r^{\uparrow\downarrow}\gamma^2 h^2 \hbar \left\{4\pi M_S \gamma + \sqrt{(4\pi M_S)^2 \gamma^2 + 4\omega^2}\right\}}{8\pi\alpha^2\{(4\pi M_S)^2\gamma^2 + 4\omega^2\}}, \quad (3)$$

where $g_r^{\uparrow\downarrow}$, $\gamma$, $h$, $\hbar$, $M_s$ and $\alpha$ are the real part of the mixing conductance, the gyromagnetic ratio of Py, the microwave magnetic field, the Dirac constant, the saturation magnetization and the Gilbert damping constant, respectively. $\omega(=2\pi f)$ is the angular frequency of the magnetization precession, where $f$ is the microwave frequency. $g_r^{\uparrow\downarrow}$ is given by [10]

$$g_r^{\uparrow\downarrow} = \frac{2\sqrt{3}\pi M_S \gamma d_F}{g\mu_B \omega}(W_{Py/GaAs} - W_{Py}), \quad (4)$$

where $d_F$, $g$, $\mu_B$, $W_{Py/GaAs}$, and $W_{Py}$ are the thickness of the Py layer, the $g$ factor, the Bohr magneton, and the linewidth of the FMR spectrum for the Py layer with and without $n$-GaAs/Pt, respectively. In this study, $d_F$, $h$, $W_{Py/GaAs}$, and $W_{Py}$ are 25 nm, 0.06 mT ( at $P_{MW}$=200 mW ), 2.97 mT, and 2.78 mT, respectively. Using Eq. (4), $g_r^{\uparrow\downarrow}$ at the Py/$n$-GaAs interface was calculated to be 6.72×10$^{18}$ m$^{-2}$. $J_s^0$ was calculated using Eq. (3) with $g_r^{\uparrow\downarrow} = 6.72 \times 10^{18}$ m$^{-2}$, $M_S = 0.0792$ T, $\gamma = 1.84 \times 10^{11}$ T$^{-1}$s$^{-1}$, $\omega = 6.05 \times 10^{10}$ rads$^{-1}$, and $\alpha = 0.00823$, $J_s^0$ to be $4.27 \times 10^{-10}$ Jm$^{-2}$.

The finite element method (MatLab R2014a PEDtool) was used for the calculation of spin current density in the GaAs channel and the Pt electrode.[34] A schematic illustration of the device structure for the calculation is shown in Fig. 3(a). In the spin diffusion model, the spin density $S$ is expressed as

$$\frac{\partial S}{\partial t} = D\nabla^2 S - \frac{S}{\tau}, \quad (5)$$

where $t$ is the time, $D$ is the spin diffusion coefficient, and $\tau$ is the spin lifetime. The

Neumann boundary condition was employed and it was assumed that the spin current ($J_s$ = 4.27×10$^{-10}$ J m$^{-2}$) flows through the boundary adjacent to the Py spin injector. $D$ for $n$-GaAs and Pt were set to 145 and 7.70 cm$^2$ s$^{-1}$, respectively, and $\tau$ in Pt $\tau_{s\_Pt}$ was set to 7.75×10$^{-14}$ s. [15] The calculated spin currents along the x and z directions, $J_{sx}$ and $J_{sz}$, between the Py and Pt electrodes are shown in Fig. 3(b). $\tau$ in the GaAs channel $\tau_{s\_GaAs}$ was set to 10, 100, and 1000 ps. The spin current beneath the Py spin injector flows in the -z direction, whereas the spin current near the edge of the spin injector flows in the +x direction. Beneath the Pt electrode, the spin current in the +z direction is enhanced due to the strong spin absorption of the Pt electrode. Thus, $V_{ISHE}$ detected in the SPIST measurements includes the electromotive force in the GaAs layer beneath the Py spin injector ($V_{ISHE1}$) and that beneath the Pt electrode ($V_{ISHE2}$), in addition to the conventionally expected $V_{ISHE}$ in the Pt electrode ($V_{ISHE\_Pt}$). $V_{ISHE1}$ is generated at a considerable distance from the Pt electrode; therefore, the contribution of $V_{ISHE1}$ to the detected $V_{ISHE}$ was investigated.

For this estimation, an experiment was implemented using the device schematically illustrated in Fig. 4(a) to measure the dependence of nonlocal voltage ($V_{Pt}$) on the local voltage ($V_{GaAs}$). A DC voltage $V_{GaAs}$ was applied between the edges of the Py electrodes and a DC voltage between the edges of the Pt electrodes $V_{Pt}$ was measured. The 10% of $V_{GaAs}$ was detected in $V_{Pt}$, as shown in Fig. 4(b), indicating that 10% of $V_{ISHE1}$ can be included in the observed $V_{ISHE}$. The spatially averaged spin current density along the z direction in the GaAs channel ($J_{s1,2}$) and in the Pt electrode ($J_s$) were calculated from the results presented in Fig. 3(b). The spin current density in the GaAs channel beneath the Py spin injector ($J_{s1}$) and beneath the Pt electrode ($J_{s2}$) were separately calculated, as schematically shown in Fig. 5(a). Figures 5(b) and (c) show $J_{s1}$, and $J_{s2}$ and $J_s$ as functions of the spin diffusion length in $n$-GaAs, $\lambda_s$, respectively. The flow direction of $J_{s1}$ is the -z direction. The magnitude of $J_{s1}$ increases sublinearly with $\lambda_s$. In contrast, $J_s$ and $J_{s2}$ have positive values and increase superlinearly with $\lambda_s$. Importantly, the ratio of $J_{s2}/J_s$ yields 1.9~2.2 at $\lambda_s \leqq$ 1.5 μm, indicating that contribution of $J_{s2}$ is non-negligible when $\theta_{SHE}$ of the channel is comparable with that of the spin detector. Figure 5(c) shows the calculated $V_{ISHE2}$, $V_{ISHE\_Pt}$, and total $V_{ISHE}$ (= $V_{ISHE\_Pt}$+$V_{ISHE2}$+0.1×$V_{ISHE1}$) as a function of $\lambda_s$. Here, $\theta_{SHE}$ of Pt was set to 0.10.[15] From the experimental $V_{ISHE}$ ( = 2.42 μV mm$^{-1}$), $\lambda_s$ and $\tau_{s\_GaAs}$ in the $n$-GaAs layer are calculated to be 1.09 μm and 81.9 ps, respectively. In the conventional one-dimensional spin diffusion model without consideration of the SOI in the channel,[1] $\lambda_s$ and $\tau_{s\_GaAs}$ were calculated to be 1.81 μm and 227 ps, of which the $\lambda_s$ obtained is approximately twice that from two-dimensional spin diffusion model with consideration of the ISHE of GaAs. Considering the previous study with an electrical nonlocal 4-terminal measurement, where $\lambda_s$ at 205 K was reported to be 1.6 μm, the value obtained from the two-dimensional spin diffusion model with the SOI of the channel used in this study is reliable compared to the conventional model. The contribution of $V_{ISHE1}$ and $V_{ISHE2}$ to the obtained $V_{ISHE}$ are 9 and 20 %, respectively. The effect of $V_{ISHE1}$ and $V_{ISHE2}$ is expected to be pronounced when a spin detector with a modest $\theta_{SHE}$, e.g., Pd ($\theta_{SHE}$=0.017) is used. [10] In this case, the contribution of $V_{ISHE2}$ is estimated to be approximately 59 % when the obtained $V_{ISHE}$ is 2.42 μV mm$^{-1}$. Therefore, this study has enabled the range of application of the SPIST method for estimation of the spin transport properties to be extended.

Finally, we comment on other contributions expected in the SPIST method. Other factors which affect the obtained $V_{ISHE}$ are the spin accumulation in localized states near the Py/GaAs interface and spin-memory loss at Py/GaAs interface, [35, 36] which cause overestimation of $J_s^0$ obtained from FMR line width. It is expected that investigation of $V_{ISHE}$ as a function of distance between spin injector and detector $L_{ID}$ reveals more accurate spin-diffusion properties and contribution of such effects, separately. An improved calculation method based on two dimensional spin diffusion model which can be used for analysis of $L_{ID}$ dependence of $V_{ISHE}$ is desired.

In conclusion, we have successfully demonstrated spin-transport in the $n$-GaAs channel at RT using the SPIST method. The electromotive force detected by the Pt spin detector has opposite polarity to that measured in the Py/GaAs bilayer structure, which demonstrates spin transport in the $n$-GaAs channel. A two-dimensional spin diffusion model that considers the ISHE of the GaAs channel was established, which revealed the accurate spin-diffusion length of $n$-GaAs ($\lambda_s$ = 1.09 μm).

### Acknowledgements
This research was supported in part by a Grant-in-Aid for Scientific Research from the Ministry of Education, Culture, Sports, Science and Technology (MEXT) of Japan (Innovative Area "Nano Spin Conversion Science" KAKENHI No. 26103003).

### Additional information
The authors declare no competing financial interests.

# References

[1] E. Shikoh, K. Ando, K. Kubo, E. Saitoh, T. Shinjo, and M. Shiraishi, Phys Rev. Lett. **110**, 127201 (2013).

[2] Z. Y. Tang, E. Shikoh, H. Ago, K. Kawahara, Y. Ando, T. Shinjo, and M. Shiraishi, Phys. Rev. B **87**, 140401(R) (2013).

[3] Y. Kitamura, E. Shikoh, Y. Ando, T. Shinjo, and M. Shiraishi, Scientific Reports **3**, 1739 (2013).

[4] S. Watanabe, K. Ando, K. Kang, S. Mooser, Y. Vaynzof, H. Kurebayashi, E. Saitoh, and H. Sirringhaus, Nature Physics **10**, 308 (2014).

[5] S. Mizukami, Y. Ando, and T. Miyazaki, Phys. Rev. B **66**, 104413 (2002).

[6] Y. Tserkovnyak, A. Brataas, and G. E. W. Bauer, Phys Rev. Lett. **88**, 117601 (2002).

[7] A. Brataas, Y. Tserkovnyak, G. E. W. Bauer, and B. I. Halperin, Phys Rev. B **66**, 060404(R) (2002).

[8] M. V. Costache, M. Sladkov, S. M. Watts, C. H. van der Wal, and B. J. van Wees, Phys Rev. Lett. **97**, 216603 (2006).

[9] E. Saitoh, M. Ueda, H. Miyajima, and G. Tatara, Appl. Phys. Lett. **88**, 182509 (2006).

[10] K. Ando and E. Saitoh, J. Appl. Phys. **108**, 113925 (2010).

[11] O. Mosendz, J. E. Pearson, F. Y. Fradin, G. E. W. Bauer, S. D. Bader, and A. Hoffmann, Phys Rev. Lett. **104**, 046601 (2010).

[12] A. Azevedo, L. H. Vilela-Leão, R. L. Rodríguez-Suárez, A. F. Lacerda Santos, and S. M. Rezende, Phys. Rev. B **83**, 144402 (2011).

[13] Y. Ando, K. Ichiba, S. Yamada, E. Shikoh, T. Shinjo, K. Hamaya, and M. Shiraishi, Phys. Rev. B **88**, 140406(R) (2013).

[14] Z. Tang, Y. Kitamura, E. Shikoh, Y. Ando, T. Shinjo, and M. Shiraishi, Appl. Phys. Express **6**, 083001 (2013).

[15] H. L. Wang, C. H. Du, Y. Pu, R. Adur, P. C. Hammel, and F. Y. Yang, Phys. Rev. Lett. **112**, 197201 (2014).

[16] F. J. Jedema, A. T. Filip and B. J. van Wees, Nature **410**, 345(2001).

[17] F. J. Jedema, H. B. Heersche, A. T. Filip, J. J. A. Baselmans and B. J. van Wees, Nature **416**, 713(2002).

[18] N. Kuhlmann, C. Swoboda, A. Vogel, T. Matsuyama, and G. Meier, Physical Review B **87**, 104409 (2013).

[19] F. J. Morin and J. P. Maita, Phys. Rev. **96**, 28(1954).

[20] A. V. Kimel, F. Bentivegna, V. N. Gridnev, V. V. Pavlov, R. V. Pisarev, and Th. Rasing Phys. Rev. B **63**, 235201 (2001).

[21] Y. K. Kato, R. C. Myers, A. C. Gossard, and D. D. Awschalom, SIENCE **306**, 1910

(2004).

[22] S. Matsuzaka, S., Ohno, Y. & Ohno, H. Phys. Rev. B **80**, 241305(R) (2009).

[23] X. Lou, C. Adelmann, S. A. Crooker, E. S. Garlid, J. Zhang, K. S. M. Reddy, S. D. Flexner, C. J. Palmstrøm, and P. A. Crowell, Nat. Phys.**3**, 197 (2007).

[24] E. S. Garlid, Q. O. Hu, M. K. Chan, C. J. Palmstrøm, and P. A. Crowell, Phys Rev. Lett. **105**, 156602 (2010).

[25] G. Salis, A. Fuhrer, R. R. Schlitter, L. Gross,and S. F. Alvarado, Phys. Rev. B **81**, 205323 (2010).

[26] K. Ando, S. Takahashi, J. Ieda, H. Kurebayashi, T. Trypiniotis, C. H. W. Barnes, S. Meakawa, and E. Saitoh, Nature Mater. **10**, 655-659 (2011).

[27] T. Uemura, T. Akiho, M. Harada, K. Matsuda, and M. Yamamoto, Appl. Phys. Lett. **99**, 082108 (2011).

[28] T. Akiho, J. Shan, H. Liu, K. Matsuda, M. Yamamoto, T. Uemura, Phys. Rev. B **87**, 235205 (2013).

[29] Y. Ebina, T. Akiho, H. Liu, M. Yamamoto, and T. Uemura, Appl. Phys. Lett. **104**, 172405 (2014).

[30] J. Shiogai, M. Ciorga, M. Utz, D. Schuh, M. Kohda, D. Bougeard, T. Nojima, J. Nitta, and D. Weiss, Phys. Rev. B **89**, 081307(R) (2014).

[31] J.-C. R.-Sánchez, M. Cubukcu, A. Jain, C. Vergnaud, C. Portemont, C. Ducruet, A. Barski, A. Marty, L. Vila, J. –P. Attané, E. Augendre, G. Desfonds, S. Gambarelli, H. Jaffrés, J. –M. George, and M. Jamet, Phys. Rev. B **88**, 064403 (2013).

[32] C. H. Du, H. L. Wang, Y. Pu, T. L. Meyer, P. M. Woodward, F. Y. Yang and P. C. Hammel, Phys. Rev. Lett. **111**, 247202 (2013).

[33] K. Ando and E. Saitoh, Nat. Comm. **3**, 629 (2012).

[34] http://www.mathworks.co.jp/products/pde/

[35] M. Tran, H. Jaffrés, C. Deranlot, J. –M. George, A. Fert, A. Mlard, and A. Lemaître, Phys. Rev. Lett. **102**, 036601 (2009).

[36] J.-C. R.-Sánchez, N. Reyren, P. Laczkowski, W. Savero, J. –P. Attané, C. Ducruet, M. Jamet, J. –M. George, L. Vila, and H. Jaffrés, Phys. Rev. Lett. **112**, 106602 (2014).

**Figure Captions**

**Figure 1 (Color online)**

**(a)** Schematic illustration of the GaAs-based device for measurement of the spin-pumping-induced spin-transport. **(b)** FMR spectra, $dI(H)/dH$, at $\theta_H = 0°$, 90°, and 180° as a function of $H$-$H_{FMR}$, where $I$ is the microwave absorption intensity in arbitrary units. The microwave power $P_{MW}$ was 200 mW. The $H_{FMR}$ at $\theta_H = 0°$ was 102 mT. **(c)** Electromotive force $V_{EMF}/w$ measured at the Pt spin detector with $\theta_H = 0°$, 90°, and 180°. The open circles represent experimental data and the solid line is a fit obtained using Eq. (1) with consideration of the contributions from the ISHE and AHE. The dashed and dotted lines are fits for the $V_{ISHE}$ signal from the Pt layer and the $V_{AHE}$ signal from the Py layer, respectively. **(d)** Magnitude of $V_{ISHE}$ as a function of $P_{MW}$.

**Figure 2 (Color online)**

**(a)** Schematic illustration of the $Ni_{80}Fe_{20}$/GaAs bilayer device. **(b)** FMR spectra, $dI(H)/dH$, at $\theta_H = 0°$, 90°, and 180° as a function of $H$-$H_{FMR}$ measured at $P_{MW} = 200$ mW. **(c)** $V_{EMF}/w$ measured between the Au/Cr contacts at $\theta_H = 0°$, 90°, and 180°. The open circles represent experimental data and the solid line is a fit obtained using Eq. (1) with consideration of the contributions from the ISHE and AHE. The dashed and dotted lines are fits for the ISHE signal from the $n$-GaAs layer and the AHE signal from the Py layer, respectively.

**Figure 3**

(a) Schematic illustration of the GaAs-based device for calculation of the two-dimensional spin diffusion model. (b) Color maps of the calculated spin current density along the x ($J_{sx}$) and z ($J_{sz}$) directions. The spin lifetime in GaAs $\tau_{s\_GaAs}$ was set to 10, 100, 1000 ps, respectively.

**Figure 4 (Color online)**

(a) Schematic illustration of the GaAs-based device for local and nonlocal voltage measurements. (b) Local voltage $V_{Pt}$ as a function of nonlocal voltage $V_{GaAs}$ measured at room temperature.

**Figure 5 (Color online)**

(a) Schematic illustration of the spin current density in the GaAs channel beneath the Py spin injector ($J_{s1}$) and beneath the Pt electrode ($J_{s2}$). (b) $J_{s1}$ and (c) $J_{s2}$ and $J_s$ as a function of spin diffusion length, $\lambda_s$ in GaAs. The inset shows the ratio of $J_{s2}/J_s$ as a function of $\lambda_s$. (d) $V_{ISHE}$, $V_{ISHE\_Pt}$, and $V_{ISHE2}$ as a function of $\lambda_s$.

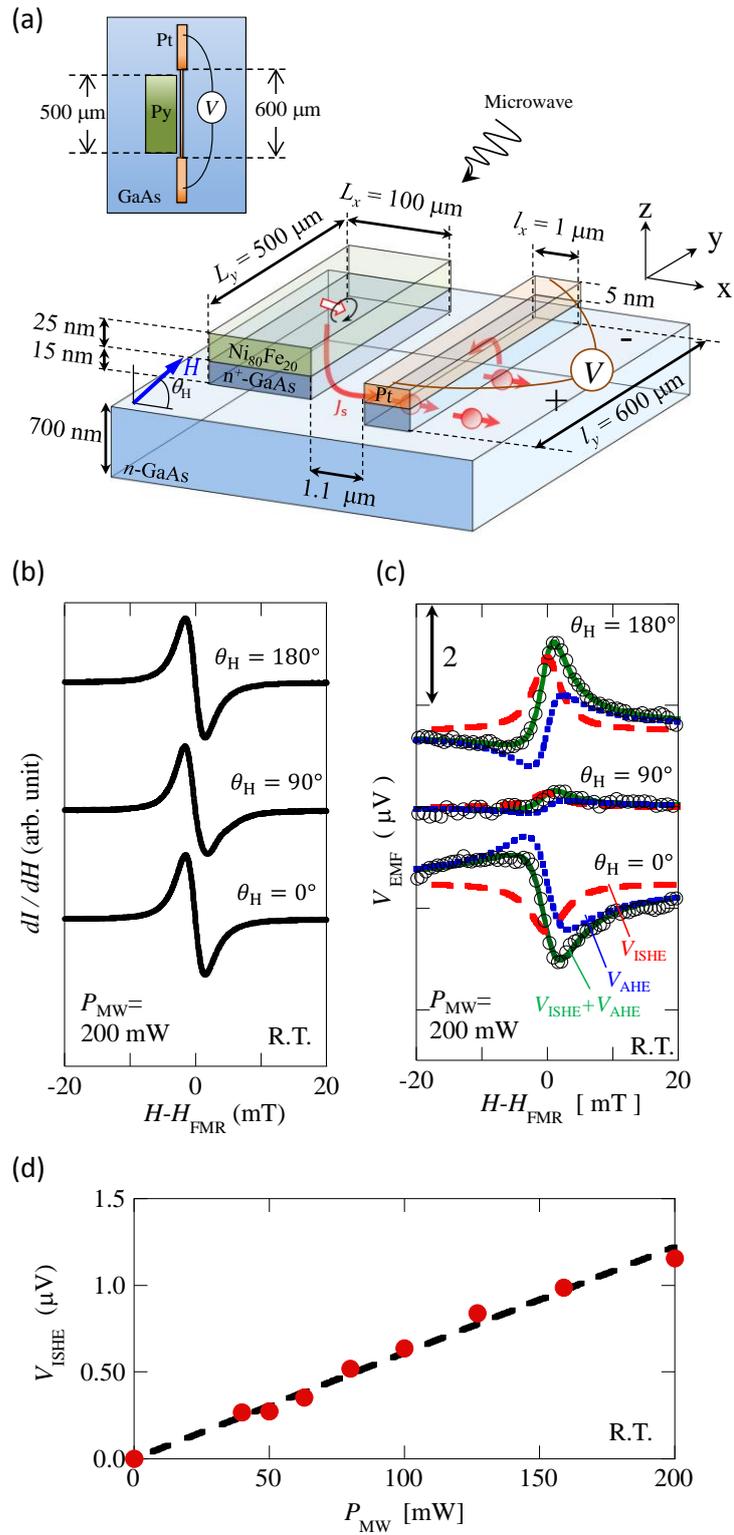

Fig. 1 A. Yamamoto et al

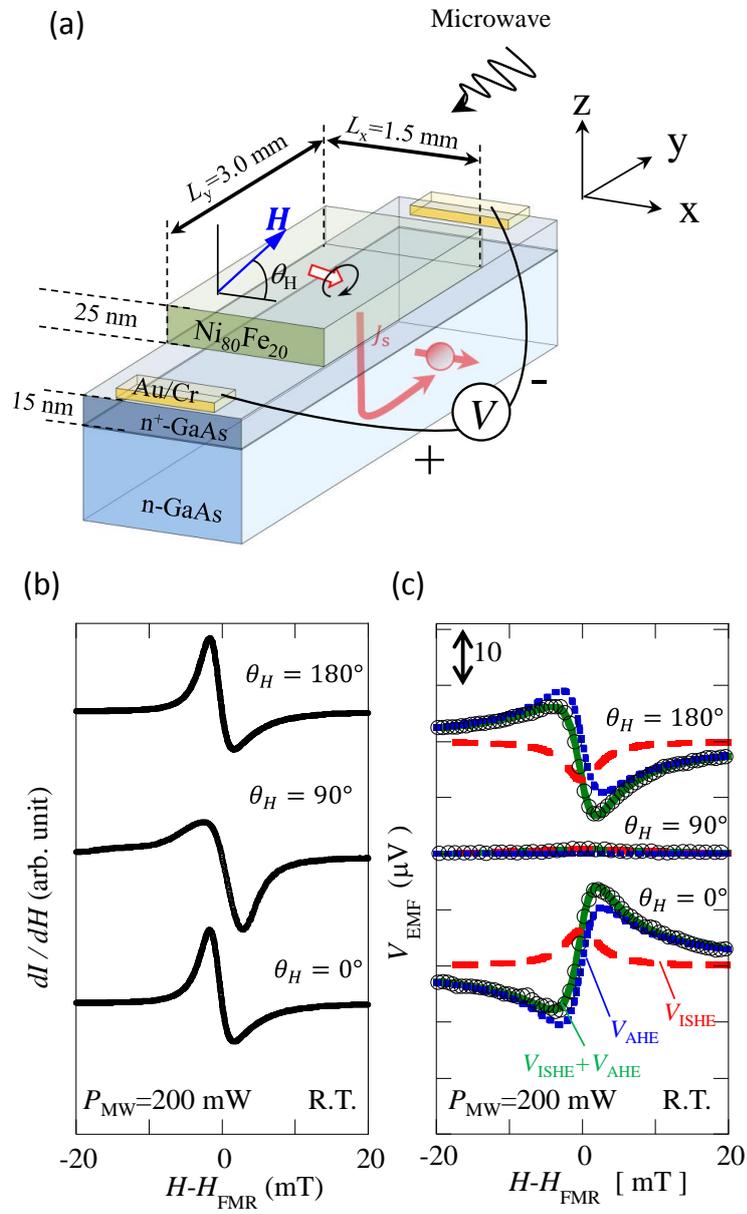

Fig. 2 A. Yamamoto et al

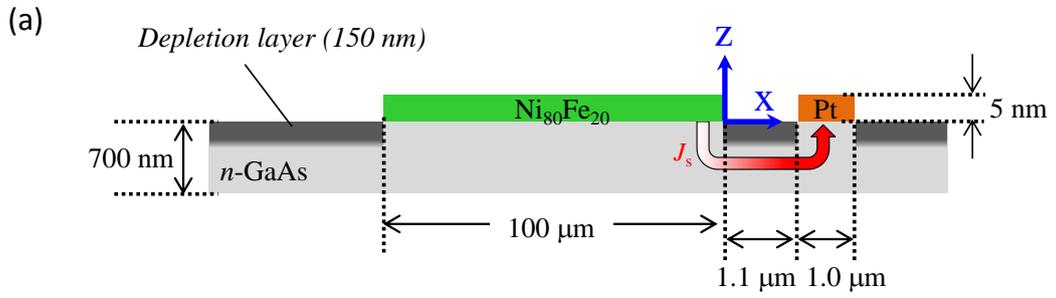

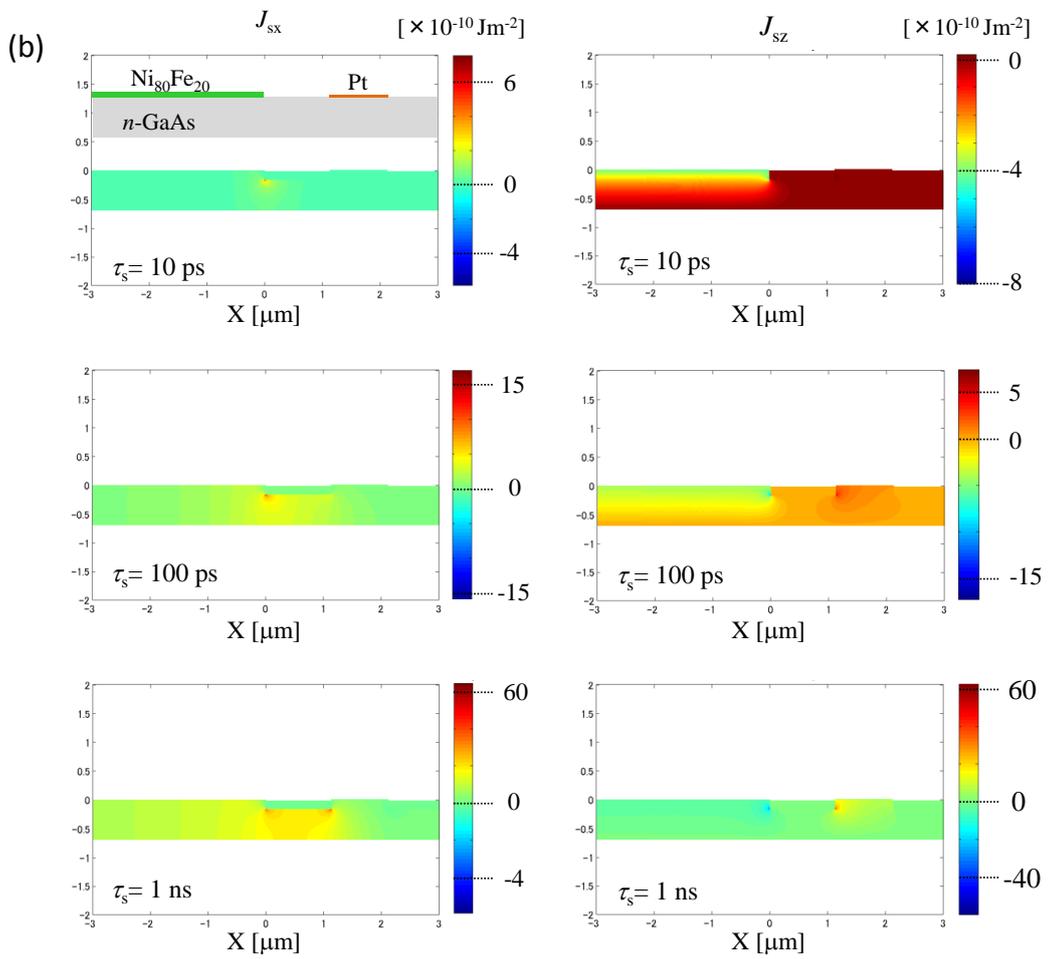

Fig. 3 A. Yamamoto et al

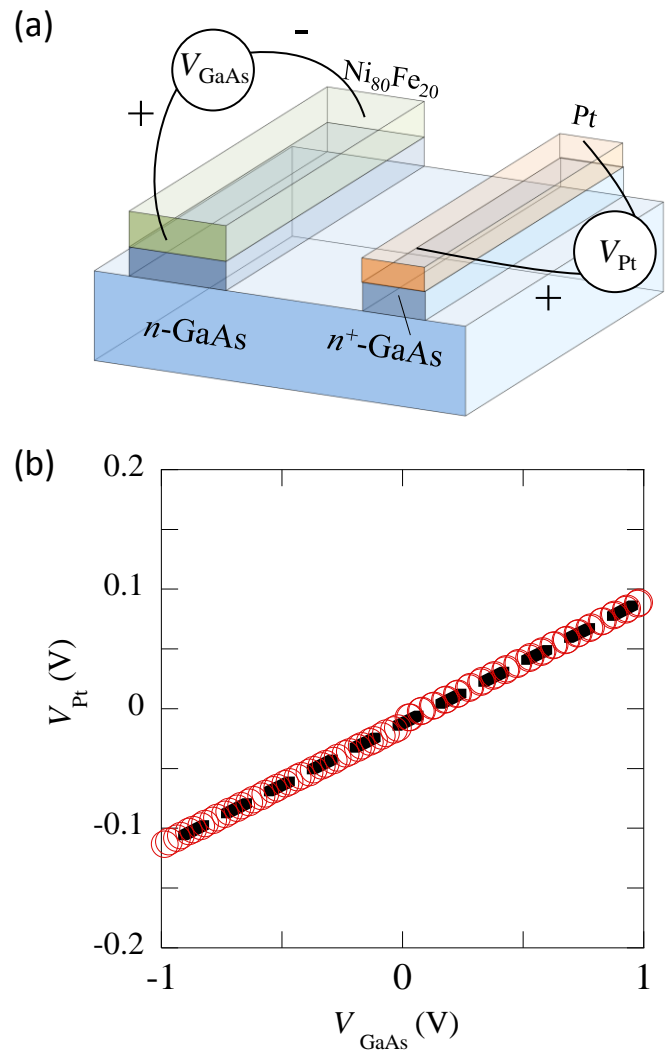

Fig. 4 A. Yamamoto et al

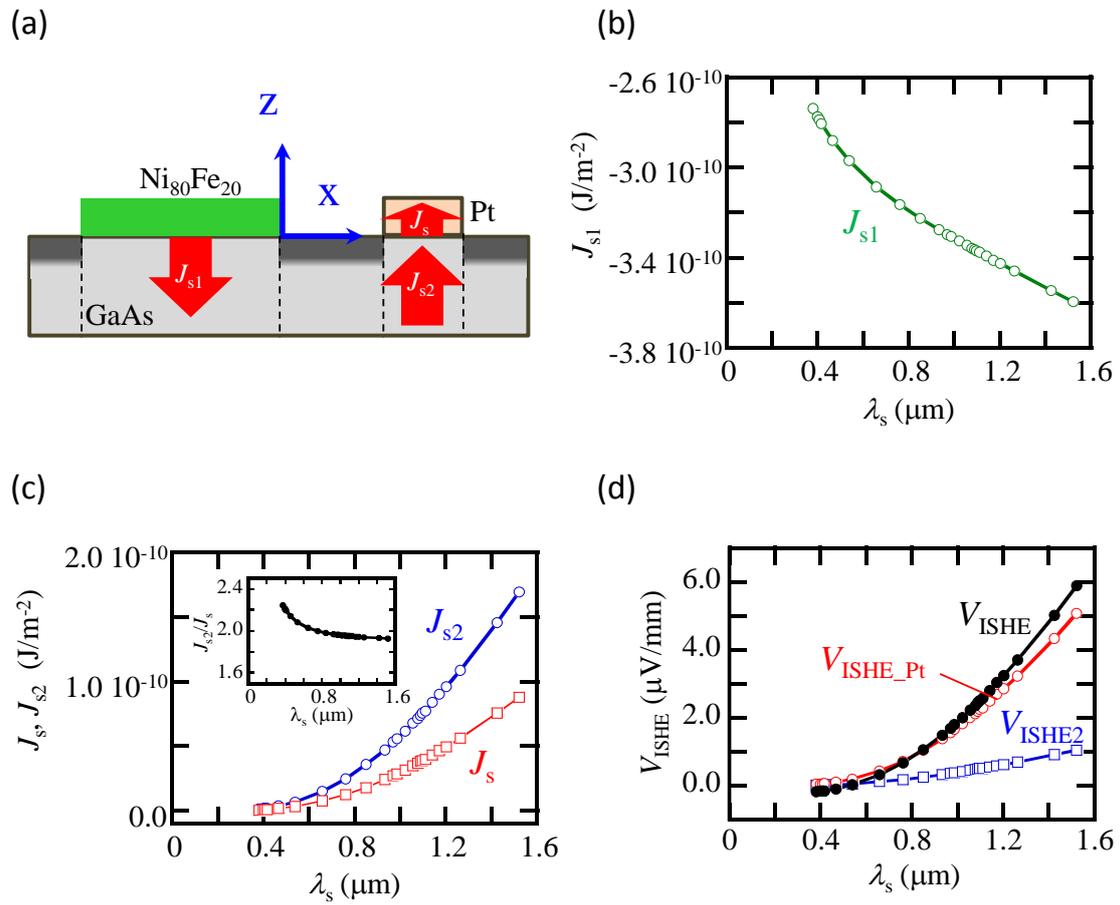

Fig. 5 A. Yamamoto et al